\documentclass[aps,prb,reprint,groupedaddress,amssymb,superscriptaddress]{revtex4-1}
\usepackage[pdftex]{graphicx}
\usepackage{bm}      
\usepackage[normalem]{ulem}    
\newcommand{\old}[1]{ } 

\newcommand{\ybcoy}{Y\-Ba$_2$\-Cu$_3$\-O$_{6+y}$}

\newcommand{\calabalacuo}{Ca$_{x}$\-La$_{1.25}$\-Ba$_{1.75-x}$\-Cu$_3$\-O$_{6+y}$}

\newcommand{\ycabcoy}{Y$_{1-x}$Ca$_x$\-Ba$_2$\-Cu$_3$\-O$_{6+y}$}

\newcommand{\ycabco}{Y$_{1-x}$Ca$_x$\-Ba$_2$\-Cu$_3$\-O$_{6}$}
\newcommand{\lsco}{La$_{2-x}$Sr$_x$\-CuO$_4$}

\newcommand{\msr}{$\mu$SR}

\begin{document}


\title{Competing orders suppressed by disorder around a hidden quantum critical point in cuprate high $T_c$ superconductors.}


\author{S.~Sanna}
\email[]{Samuele.Sanna@unipv.it}
\affiliation{Dipartimento di Fisica A.~Volta and CNISM, Universit\`a di Pavia, Via Bassi, 6, I-27100 Pavia, Italy}
\author{F.~Coneri}
\affiliation{Dipartimento di Fisica, Universit\`a di Parma, Viale G.Usberti, 7A, I-43100 Parma, Italy}
\author{A.~Rigoldi}
\affiliation{Dipartimento di Fisica, Universit\`a di Cagliari, Italy}
\author{G.~Concas}
\affiliation{Dipartimento di Fisica, Universit\`a di Cagliari, Italy}
\author{S.~R.~Giblin}
\affiliation{ISIS Facility, STFC-Rutherford Appleton Lab, HSIC, Didcot, OX110QX, U. K.}
\author{R.~De Renzi}
\affiliation{Dipartimento di Fisica and CNISM, Universit\`a di Parma, Viale G.Usberti, 7A, I-43100 Parma, Italy}




\begin{abstract}
We report extensive muon spin rotation measurements on the lightly doped  \ycabcoy\ compound, which allows us to disentangle the effect of disorder, controlled by random Ca$^{2+}$ substitution, from that of mere doping.
A 3D phase diagram of lightly doped cuprates is accurately drawn. It shows a quantum critical point around which a {\em thermally activated} antiferromagnetic phase competes with
superconductivity. Disorder suppresses both the competing order parameters and the quantum critical point, unveiling an underlying frozen state.
\end{abstract}


\maketitle


It is well known that the high $T_c$ superconductivity of cuprates occurs nearby the suppression of an antiferromagnetic (AF) phase. In the pure \ybcoy\  compound \cite{TranquadaPRB1989,Rossat-MignodPB1990,AlloulPRL1991,SannaPRL2004,AlloulRMP2009} for increasing hole doping the AF phase is directly followed by the onset of superconductivity (SC). The incipient superconductor coexists with low temperature magnetic order, \cite{SannaPRL2004} often called cluster spin glass. In \ycabco, \lsco\ and \calabalacuo\ the latter becomes the sole order parameter in a wide doping range \cite{NiedermayerPRL1998,KerenPRB2006} between the suppression of $T_N$ and the onset of $T_c$. These systems are classified as dirty \cite{DagottoScience2005} since they contain disordered substitutional heterovalent cations, i.e. strong Coulomb impurities in the vicinity of the CuO$_2$ layers. In this respect pure \ybcoy\ approaches the clean limit.

\begin{figure}
\includegraphics[width=0.45\textwidth,trim= 5 2 20 25, clip,angle=0]{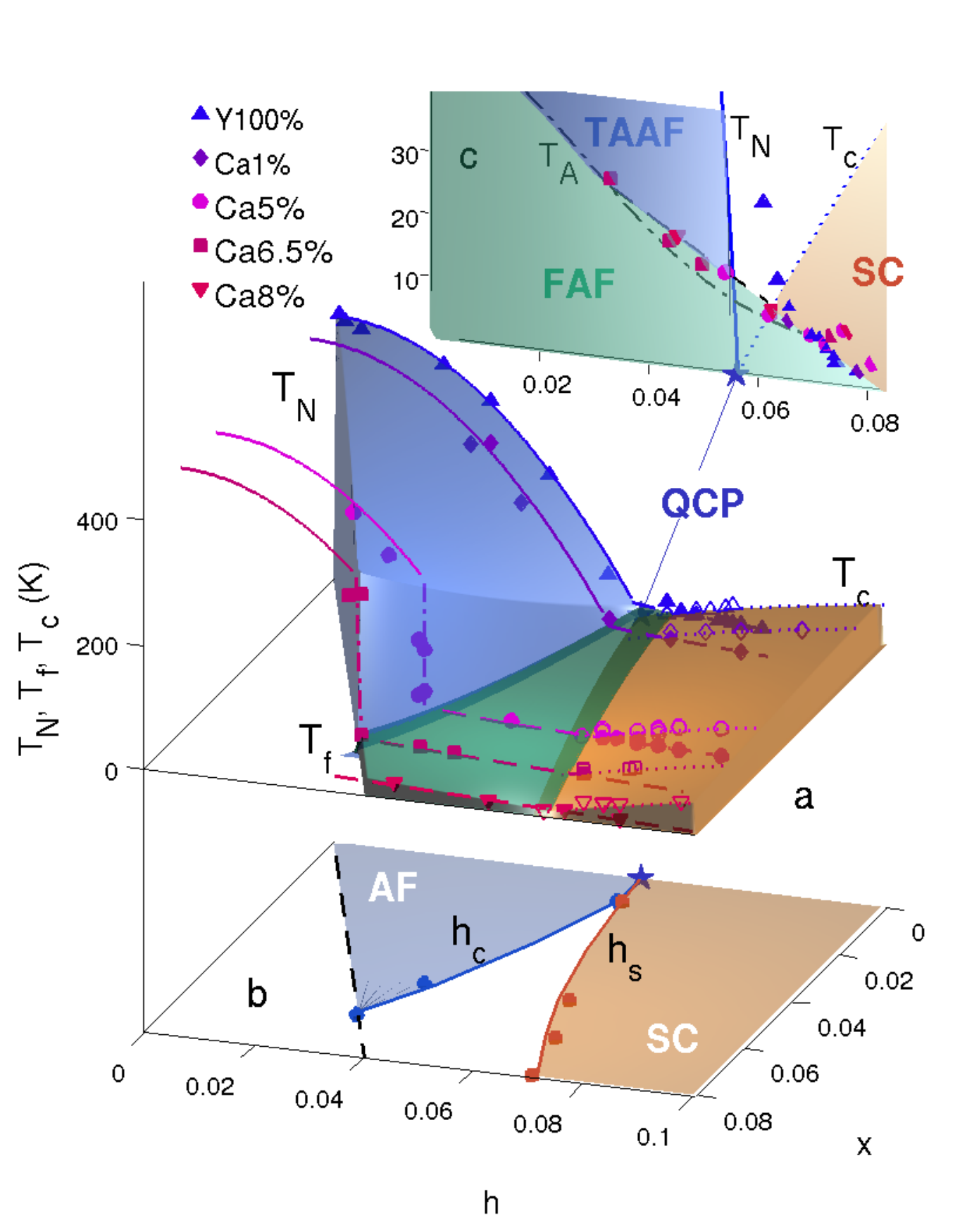}%

\caption{\label{fig:3D} (color on-line) Doping ($h$) and disorder ($x$, Ca content) dependence in \ycabcoy: a) Phase diagram; b) $x,h$ projection with critical hole densities $h_c(x)$, $h_s(x)$, and lower physical limit ($h=x/2$, dashed); c) Rescaled projection of $T_f(h,x)$ on the $x=0$ plane; solid, dotted and dashed-dotted curves are the clean limit  $T_N$, $T_c$, and $T_A$, respectively, from Ref.~\onlinecite{ConeriPRB2010}.}
\end{figure}

The lightly doped AF phase of both clean and dirty cuprates is characterized by two regimes: \cite{ChouPRL1993,NiedermayerPRL1998,SannaSSC2003} a low temperature re-entrant phase, where the Cu moments recover the full value of 0.6 $\mu_B$ appropriate for a two dimensional Heisenberg quantum antiferromagnet, 
and a high temperature antiferromagnetic phase, where both the moment $m(h)$ and $T_N(h)$ are strongly reduced by the hole doping $h$.
We clarified the {\em clean limit} behavior in a recent work \cite{ConeriPRB2010} demonstrating the following features: The crossover between the two regimes is connected  with a low temperature thermal activation of charge carriers; $T_N$ is actually the critical temperature of a {\em thermally activated} antiferromagnet (dubbed TAAF), characterized by a metallic behavior; \cite{AndoPRL2001} The magnetic ground states in the re-entrant region and, at higher doping, in the coexistence region are the same {\em frozen} antiferromagnet (dubbed FAF); The curve describing $T_N(h)$ and $T_c(h)$ merge at $T=0$ for $h_{c0}=0.056(2)$, suggesting the presence of a quantum critical point (QCP) hidden underneath (i.e. replaced by) the FAF state. The three regimes, TAAF, FAF and SC, and their clean limit crossover \footnote{\protect{$T_A(h)$ is obtained in Ref.~\onlinecite{ConeriPRB2010} from the fit of a phenomenological thermally activated model for the local magnetic moment seen by the muon.}} $T_A(h)$ are shown in Fig.~\ref{fig:3D}c.

Here we report an accurate investigation of the cuprate behavior under controlled disorder conditions. Our results show that the competing AF and SC orders are suppressed around the hidden QCP by Coulomb-driven disorder, which unveils the underlying intrinsic FAF state, that also coexists with superconductivity. We investigated the model system \ycabcoy\ where the random Coulomb potential of Ca$^{2+}$ impurities controls disorder, while hole doping is tuned by oxygen content. Five series of polycrystalline samples are investigated, with $x=0,0.01,0.05,0.065,0.08$ (henceforth Y100\%, Ca1\%, Ca5\%, Ca6.5\%, Ca8\%), which are the same as in Ref.~\onlinecite{SannaPRB2008}.

Figure \ref{fig:3D}a shows a three dimensional view of the \ycabcoy\ phase diagram, that summarizes our results. 
The data are obtained as follows:  the N\'eel $T_N$ and the low temperature order transition $T_f$, from Zero-Field (ZF) \msr, applying standard analysis; \cite{SannaSSC2003,ConeriPRB2010} the superconducting $T_c$ from susceptibility and the hole content $h$ from thermopower. \cite{SannaPRB2008}
These results disentangle the influence of doping, $h$, and disorder, $x$, showing that the green region of Fig.~\ref{fig:3D}a where $T_f$ is the only transition continuously widens with increasing Ca content, i.e. with disorder. A similar conclusion was qualitatively shown in earlier work on Zn substitution, \cite{AlloulPRL1991} across different cuprate families, \cite{NiedermayerPRL1998,SannaJSNM2005} and in irradiated samples. \cite{RullierEPL2008}

Let us clarify the distinct influence of doping and disorder on Fig.~\ref{fig:3D}a. Dotted parabolae describe the superconducting transitions $T_c(h)$, which shift rigidly to higher onsets $h_s(x)$ with Ca content, as previously reported.\cite{SannaPRB2008} Solid parabolae represent the doping dependence of the N\'eel temperature $T_N(h)=T_{N0}\left[1-(h/h_{c0})^2\right]$  in the clean limit, with $T_{N0}=422(5)$ K and $h_{c0}=0.056(2)$ \cite{ConeriPRB2010}.
Remarkably the same function, with no adjustable parameters, agrees with the low hole density data of Ca1\% (diamonds) and Ca5\% (circles), demonstrating that $T_N(h)$ is, at least initially, a universal function of doping, independent of the disorder parameter $x$. This behavior is interrupted at Ca dependent critical hole densities, $h_c(x)$, by a first-order transition from high $T_N$ values to much lower $T_f$ values (vertical dash-dotted lines in Fig.~\ref{fig:3D}a). A similar abrupt jump is displayed also by \calabalacuo, a system where doping can be varied at fixed disorder.\cite{OferPRB2008}

The two critical hole densities, $h_c(x)$ and $h_s(x)$, projected onto the $x,h$ plane in Fig.~\ref{fig:3D}b, converge at $(x,h)=(0, 0.056)$.
Hence the four curves, $T_N(0,h), T_c(0,h), h_c(x)$ and $h_s(x)$ merge at one point and the simultaneous vanishing of $T_N$ and $T_c$ defines it as a QCP, proper of the ideal cuprate. This differs from the more popular location, underneath the SC dome, although a second QCP close to the midpoint between the AF and the SC states was suggested before.\cite{PanagopoulosPRB2005,SebastianPNAS2010,SachdevPSSB2010}

\begin{figure}
\includegraphics[width=0.45\textwidth,trim= 30 210 30 290, clip,angle=0]{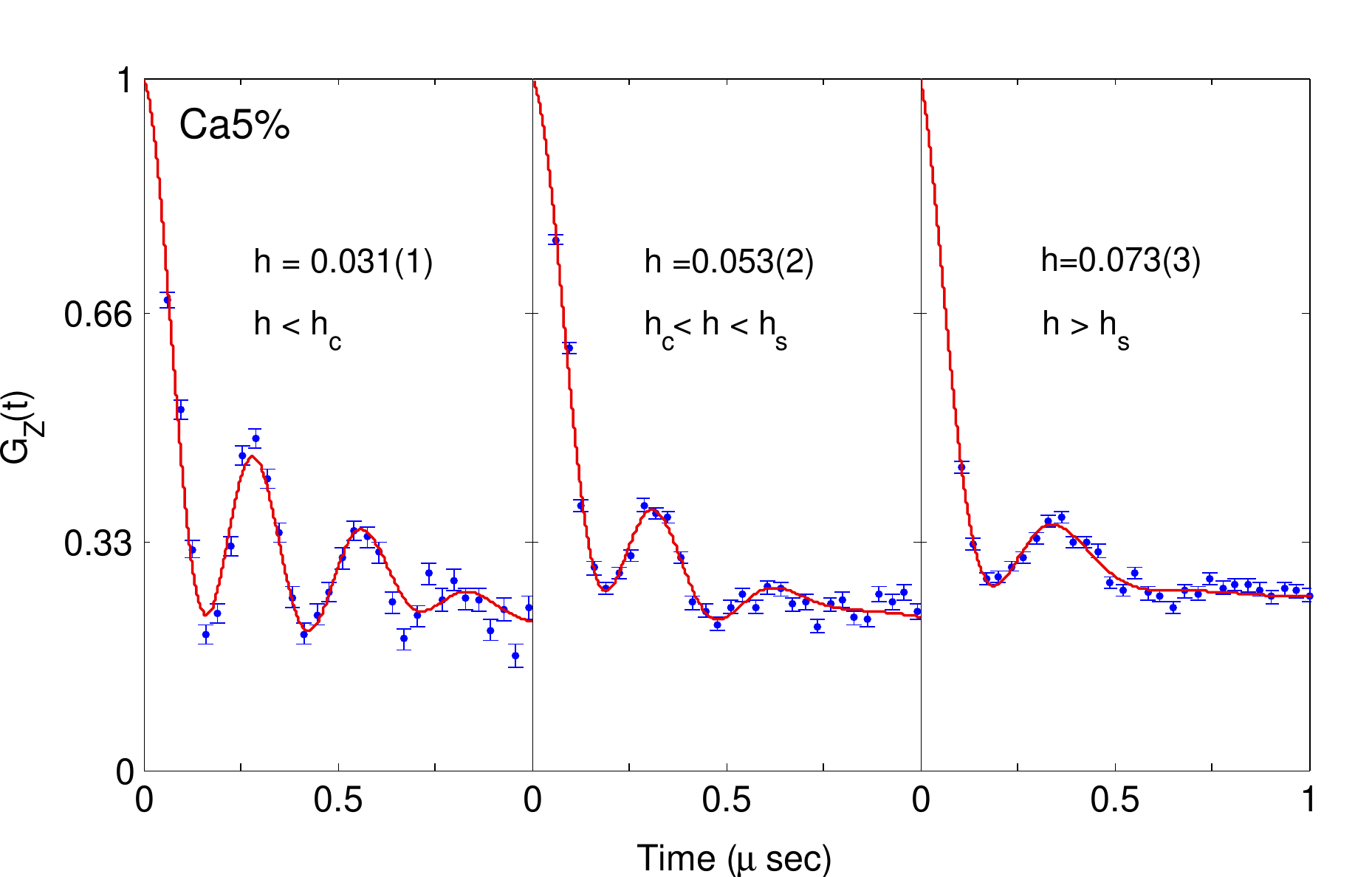}%
\caption{\label{fig:asymmetry} (color on-line)  Ca5\% ($h_c=0.04,\> h_s=0.068$): muon asymmetry at $T=2$ K with best fits for three very different hole densities. }
\end{figure}

The low transition temperature $T_f(h,x)$ (dashed curves in Fig.~\ref{fig:3D}a) are projected in Fig.~\ref{fig:3D}c onto the $(T,h)$ plane, revealing  a common doping dependence, since they all fall on the same dashed line, from $x=0$ to $x\approx 0.08$. Notice that this line lies very close to the activation temperature $T_A(h)$ detected in the clean limit compounds \cite{ConeriPRB2010} (dash-dotted curve), i.e. at the crossover between the FAF state and the thermally activated AF regime. Therefore disorder does not modify appreciably the border of the FAF phase, which is robust and hides (replaces) the QCP in pure Y100\%.

Let us now consider the mean staggered Cu moment $m(x,h,T)$ detected by  ZF \msr.
Figure \ref{fig:asymmetry} shows the very similar time evolution of the ZF muon asymmetry at $T=2$ K, in three Ca5\% samples with different hole densities, respectively $h=0.031<h_c$ ($T_N=258$ K), $h_c<h=0.053<h_s$ ($T_f=16.1$ K) and $h=0.073>h_s$ ($T_f=7.7$ K), within the green region of Fig.~\ref{fig:3D}c. Solid curves are best fits to the standard ZF functions for polycrystals.\cite{ConeriPRB2010} The oscillations are due to the muon spin precession around the magnetic field $B_\mu$ at the muon site,
\cite{ConeriPRB2010,PinkpankPhysC1999} which is proportional to the average local Cu moment. Apart from a modest frequency reduction, the main difference between the three samples is an increase of the damping, up to a maximum relative value $\Delta m/m\equiv \Delta B_\mu/B_\mu=0.3$, that indicates inhomogeneity, but still within a well ordered magnetic state, as it is demonstrated by the oscillatory pattern still present for $h>h_s$. Similar results were obtained (Fig.~10, Ref.~\onlinecite{ConeriPRB2010}) for pure Y100\%. In contrast muon precessions  in typical spin glasses are strongly overdamped, due to a very broad distribution of local fields \cite{UemuraPRB1985}, with $\Delta B_\mu/B_\mu\gtrsim 1$ (i.e.~width comparable to the mean value). As a matter of fact neutron scattering experiments\cite{StockPRB2008} on an YBa$_2$Cu$_3$O$_{6.35}$ single crystal with $h>h_s$, $Tc=18$ K  observe broad $(H H L)$ peaks with semiinteger H, integer L and an isotropic correlation length of order $\xi\approx 10\,\AA$, indicating nearly static correlations, antiferro- in the plane and ferro-magnetic among bilayers.
We propose to abandon the spin glass terminology for cuprates, recognizing that the glassy features described in earlier reports \cite{PanagopoulosPRB2005,spinglass} belong to a short range magnetic state, quite different from a conventional glass. This state is the same frozen antiferromagnet at very low doping, below $T_A$, and at higher doping and/or disorder, below $T_f$.

A crucial issue is whether AF and SC properties coexist atomically, or in nanoscopically close, but separate regions. For the $T_c\approx 18K$ samples (e.g.~$h>h_s$, third panel of Fig.~\ref{fig:asymmetry}, or equivalent Y100\% samples\cite{ConeriPRB2010}) the still large relative width $\Delta B_\mu$ indicates that some of the muon sites are not immediately surrounded by static spins. This is compatible with the presence of distinct AF and SC neighborhoods, intertwined on a length-scale of $10\,\AA$, a view that is supported by the further broadening of the width $\Delta B$, which was observed in  samples\cite{SannaPRL2004} with a higher $T_c=30$ K, where the oscillatory pattern is over-damped even at the lowest temperature.

\begin{figure}
\includegraphics[trim= 0 20 0 40, clip, width=0.45\textwidth,angle=0]{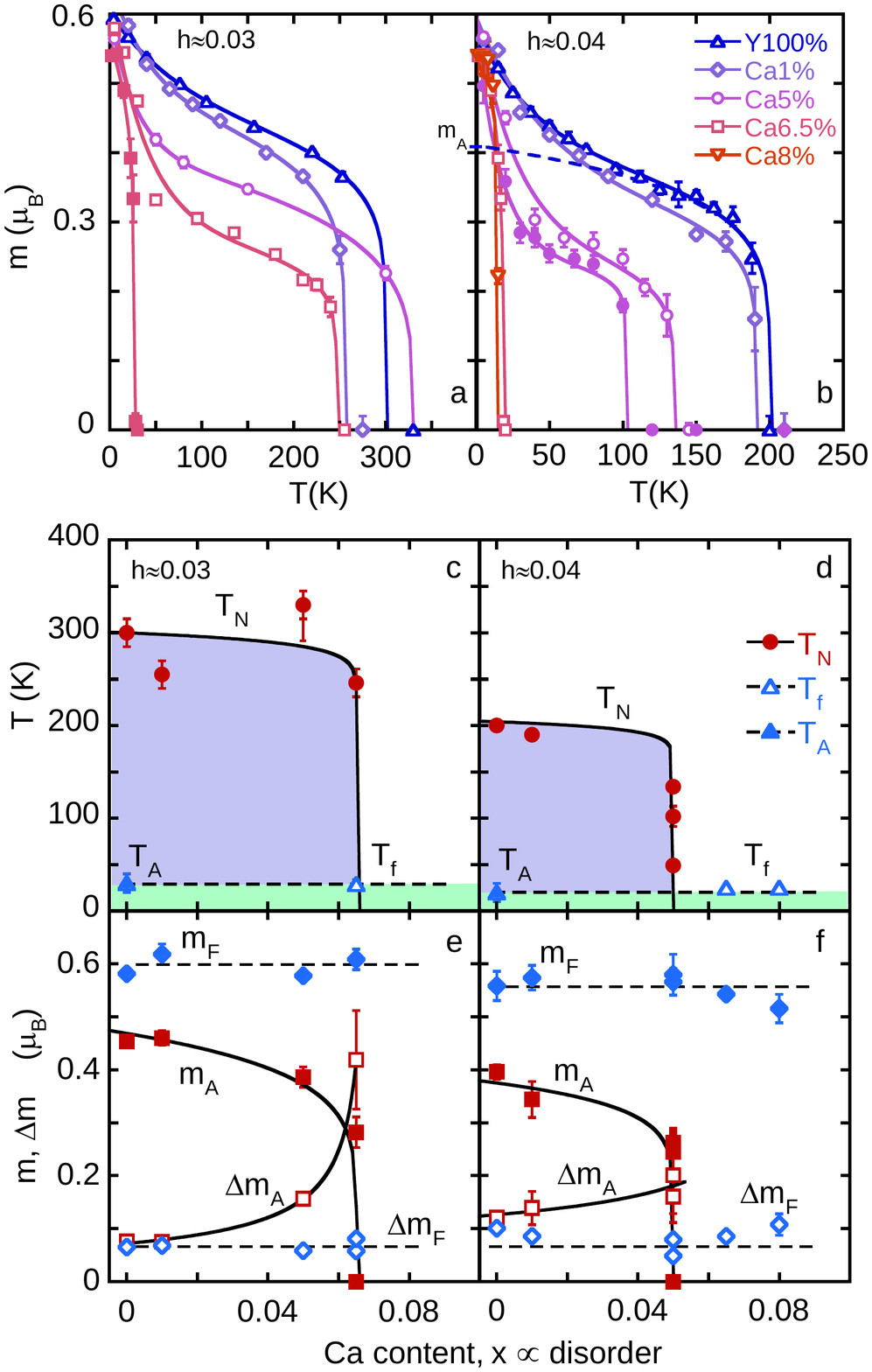}

\caption{\label{fig:fig34} (color on-line) Samples at constant hole densities: a), b) Temperature dependence of the magnetic moment with the best fit to Eq.~\ref{eq:fitc}, for the $h\approx 0.03$ and $h\approx0.04$ samples. The dashed line shows an example of the low $T$ extrapolation $m_A$; c), d) Ordering ($T_N, T_f$) and activation ($T_A$) temperatures; e), f) Magnetic moment and width of its distribution in the FAF ($m_F,\Delta m_F$) and in the TAAF regime ($m_A,\Delta m_A$).}
\end{figure}

The upper bound of the FAF region in Fig.~\ref{fig:3D}c coincides with the crossover from insulator to metal, the peculiar {\em bad metal} \cite{EmeryPRL1995} that is detected by transport \cite{AndoPRL2001,AndoJPCS2008} for all dopings at high temperature (in antiferromagnetic, paramagnetic, as well as superconducting samples). The common bad-metal character consists in a Fermi wave vector much less than the inverse mean free path, well below the Mott-Ioffe-Regel limit.\cite{AndoJPCS2008} The FAF state, of course, remains properly insulating only to the left of $h_s$, while on the right nanoscopic coexistence with percolating SC masks the intrinsic transport properties of the  FAF clusters.

We now consider the disorder dependence of the staggered magnetic moment. Since the magnetic structure does not change with doping the local muon field $B_\mu$ remains proportional \cite{ConeriPRB2010} to the staggered Cu moment. The latter is measured as $m(h,T)=m_0 B_{\mu}(h,T)/B_{\mu 0}$, where $m_0=0.6\,\mu_B$ and $B_{\mu 0}$ is the local field value in the unsubstituted parent compound at $T=0$. Let us focus on two slices of the phase diagram of Fig.~\ref{fig:3D}a, at constant doping versus disorder, for $h\approx 0.03$ and 0.04. The temperature dependence of $m(T)$ is displayed in Fig.~\ref{fig:fig34}a, b for different values of $x$ (disorder). Best-fit solid curves show that it follows the same power-law behavior discussed in Ref.~\onlinecite{SannaSSC2003,ConeriPRB2010}
\begin{equation}
\label{eq:fitc}
m(T)= [m_A + (m_F-m_A)e^{-c T}](1-T/{T_m})^\beta
\end{equation}

\noindent describing the low temperature upturn due to the smooth moment crossover between the FAF and the TAAF regimes. The parameter $m_F$ is the $T\rightarrow 0$ value of the moment in the FAF state, whereas $m_A$ is the low temperature extrapolation  of the power law in the TAAF regime, as shown e.g by the dashed line in Fig.~\ref{fig:fig34}b. Therefore they represent the $T=0$ order parameters of the FAF and TAAF phases, respectively.

The best fit parameters $m_A$ and $m_F$ are shown in Fig.~\ref{fig:fig34}, panels e, f, and the corresponding transition temperatures in panels c, d. The first order nature of the transition between samples with and without a TAAF phase is apparent also when cut along the disorder axis (panels c and d). Disorder suppresses $T_N$, and the TAAF phase with it, unveiling the underlying FAF phase  which is unaltered ($T_A\approx T_f$). The moment $m_F$ in the FAF state is constant vs.~$x$ (in each panel, e, f), and very weakly $h$-dependent (across the two panels), approaching the value $0.6\,\mu_B$ of the clean undoped parent compound. It agrees with the magnetic-site dilution regime \cite{ConeriPRB2010} appropriate for this truly insulating localized-moment state, as in clean \ybcoy. A totally different conclusion applies to the TAAF regime, where the moment $m_A$ is strongly dependent both on $x$ and $h$, due to the peculiar topology of spin and mobile holes in the TAAF, as discussed in more details in Ref.~\onlinecite{ConeriPRB2010}.
Disorder suppresses the TAAF order, apparently by amplifying the disruptive effect of thermal activation. Accordingly, the behavior of the widths is respectively disorder dependent for $\Delta m_A$ and independent for $\Delta m_F$.


\begin{figure}
\includegraphics[trim=  80 240 80 220, clip,width=0.35\textwidth,angle=0]{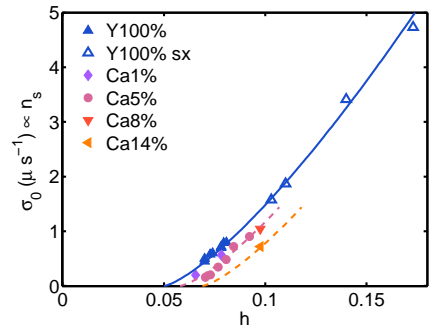}

\caption{\label{fig:ns_vs_h} (color on-line) Dependence on hole doping of the $T=0$ \msr\ relaxation rate $\sigma\propto n_s$ from TF-\msr\ data: Y100\%, filled triangle,\cite{SannaPRL2004} Y100\% crystals, open triangle,\cite{SonierPRB2007} Ca samples.\cite{SannaPRB2008} The solid curve is from fig. 8, Ref~\onlinecite{SonierPRB2007}, rigidly shifted for the Ca substitution (dashed lines).}
\end{figure}

Before concluding, we briefly recall that in the phase diagram to the right of the QCP disorder raises the threshold $h_s(x)$ for the onset of a superconducting condensate. The condensate density, $n_s$, is proportional to the
relaxation rate $\sigma_0$ measured in transverse field \msr;\cite{BarfordPC1988} Fig.~\ref{fig:ns_vs_h}
summarizes our measurements in a field of 20 mT on the same samples
investigated here.\cite{SannaPRB2008} The open symbols and the solid line represent the \ybcoy\ data and their interpolation from Ref.~\onlinecite{SonierPRB2007}, rescaled to polycrystalline values \cite{BarfordPC1988} and to low fields (fig. 6 of Ref.~\onlinecite{SonierPRB2007}). The Y100\% data fall on the same curve. For larger values of $x$ the data and their dashed interpolation shift to the right, indicating that Coulomb disorder suppresses $T_c$ and the superconducting order parameter $n_s$ with it by reducing the hole density available for superconductivity.

This is a (nearly) symmetric effect to the suppression of $T_N$ and of the related magnetic order parameter $m_A$ on the left of the QCP, where the first order transition induced by disorder precipitates all cuprates (e.g.~Fig.~\ref{fig:fig34}c, d) from the bad-metal TAAF regime into the underlying FAF insulating state, at a Ca dependent hole density $h_c(x)$. The main effect of disorder is therefore the separation of $T_N(x,h_c)$ from $T_c(x,h_s)$ (they coincide for $x=0$). The QCP is a point where both order parameters are present and compete, and their characteristic energy $kT_N,kT_c$ is reduced to zero. The QCP disappears when $T_N(x,h_c)$ and $T_c(x,h_s)$ (which coincide for $x=0$) become separated.

In the AF sector of the phase diagram the main effect of the increasing disorder appears to be a lowering of the threshold $h_c(x)$ for the disappearance of TAAF magnetism. Disorder seems to be simply unveiling the intrinsic FAF behavior, by pushing aside the metallic TAAF and SC states from the QCP at $x=0$. 
The boundary of the FAF phase in Fig.~\ref{fig:3D}c is the same for all $x$ values, at least up to $h=0.075$, indicating that most of this phase is quite insensitive to disorder. This is an argument against a cluster spin glass introduced by disorder.\cite{DagottoScience2005} The unique $T_f(h)$ and the related $T_A(h)$ rather indicate an intrinsic origin, connected to the thermally activated crossover \cite{ConeriPRB2010} in the self organized charge and spin fabric (spin spirals \cite{ShraimanPRL1989} and/or stripes \cite{ZaanenPRB1989}) of the doped Mott-Hubbard system.

Summarizing, the behavior in the whole phase diagram agrees with the following simple notions, referred to Fig.~\ref{fig:3D}c. Charge localization characterizes the green region of the phase diagram, where the insulating FAF magnetism is just diluted by doping and largely insensitive to disorder. The remaining regions correspond to a bad metal state, where the TAAF and the SC compete around a QCP. The bad metal appears by thermal activation on the left of the QCP, but it is one of the two $T=0$ ground states on the right, where SC sets in. The bad metal is very sensitive to charged impurity disorder, leading to the suppression of both its competing TAAF and SC order parameters, but not of its bad metal character, which is common \cite{AndoPRL2001} to all cuprates. We finally notice that a line of first order transitions $h_c(x)$ terminates from the left in a second order QCP. The steeper $h_s(x)$ curve represents the onset of superconductivity in the regime of (nanoscopic) phase separation, hence a corresponding first order transition line may exist farther to the right. This simple description of the interplay between disorder and doping, with two first order lines terminating at the same QCP, awaits a full theoretical justification.

\begin{acknowledgments}
This work was carried out at the ISIS muon facility (RAL). We thank Lara Benfatto, Jos\'e Lorenzana, Marco Grilli, Brian Andersen, Elbio Dagotto and Gonzalo Alvarez for stimulating discussions. Partial support of PRIN-06 project and of the EU-NMI3 Access Programme is acknowledged.

\end{acknowledgments}


\bibliography{PRhetero}

\end{document}